\documentclass[11pt]{article}
\usepackage[english]{babel}
\usepackage{amsmath}
\usepackage{amsfonts}
\usepackage{amssymb}
\begin{document}

\title{\hfill{\normalsize{}hep-th/0701243}\\[5mm]
{\bf{}Classical BRST charge for nonlinear algebras}}

\author{\sc I.L. Buchbinder${}^{a}$\thanks{joseph@tspu.edu.ru},
 P.M. Lavrov$^b$\thanks{lavrov@tspu.edu.ru}
\\[0.5cm]
\it ${}^a$Department of Theoretical Physics,\\
\it Tomsk State Pedagogical University,\\
\it Tomsk 634041, Russia\\[0.3cm]
\it ${}^b$Department of Mathematical Analysis,\\
\it Tomsk State Pedagogical University,\\
\it Tomsk 634041, Russia}
\date{}

\maketitle \thispagestyle{empty}

\begin{abstract}\noindent
We study the construction of the classical nilpotent canonical
BRST charge for the nonlinear gauge algebras where a commutator
(in terms of Poisson brackets) of the constraints is a finite
order polynomial of the constraints. Such a polynomial is
characterized by the coefficients forming a set of higher order
structure constraints.  Assuming the set of constraints to be
linearly independent, we find the restrictions on the structure
constraints  when the nilpotent BRST charge can be written in a
simple and universal form. In the case of quadratically nonlinear
algebras we find the expression for third order contribution in
the ghost fields to the BRST charge without the use of any
additional restrictions on the structure constants.

\end{abstract}

\section{Introduction}
Discovery of the BRST symmetry \cite{brs}, \cite{t} exerted a large
impact upon development of gauge theories. At present, the BRST
charge, corresponding to the Noether current of BRST symmetry, is
one of the most efficient tools for studing the classical and
quantum aspects of constrained systems (see e.g. \cite{gt}). The
properties of the BRST charge, especially its nilpotency, are the
base of modern quantization methods of gauge theores in both
Lagrangian \cite{BV} and Hamiltonian \cite{BFV} formalism (see also
the reviews \cite{BF}). Namely the nilpotency of BRST charge plays a
crucial role for generic description of physical state space in
quantum theory of gauge fields \cite{KO}. One of the most important
modern applications of the BRST charge is related to string theory
and (super)conformal theories \cite{Kato} (see also \cite{gsw})
where the use of the BRST charge allowed one to construct a physical
state space, to clarify the structure of physical spectrum and
derive the critical dimensions of the models. In addition, we point
out that the nilpotent BRST charge was used for constructing the
action of string field theory \cite{w} (see \cite{tz} for review).

In this paper we discuss the form of the canonical BRST charge (to
be more precise, the BRST - BFV charge \cite{BFV}) for a general
enough class of gauge theories. The BRST - BFV construction is based
on classical formulation of the gauge theory in phase space where
the gauge theory is characterized by first class constraints
$T_{\alpha}=T_{\alpha}(p,q)$ with $p_i$ and $q^i$ being canonically
conjugate phase variables. Constraints $T_{\alpha}$ satisfy the
involution relations in terms of the Poisson bracket
\begin{eqnarray}\label{i0}
\{T_{\alpha},T_{\beta}\}=f^{\gamma}_{\alpha\beta}T_{\gamma}
\end{eqnarray}
with structure functions $f^{\gamma}_{\alpha\beta}(q,p)$. In
Yang-Mills type theories the structure functions are constants and
the nilpotent BRST charge ${\cal Q}$ ($\{{\cal Q},{\cal Q}\}=0$) can
be written in a closed form as follows
\begin{eqnarray}\label{i1}
{\cal Q}=c^{\alpha}T_{\alpha}-\frac{1}{2}c^{\alpha}c^{\beta}
f^{\gamma}_{\alpha\beta}{\cal P}_{\gamma}
\end{eqnarray}
where $c^{\alpha}$ and ${\cal P}_{\alpha}$ are canonically
conjugate ghost variables. For general gauge theories the
structure functions depend on phase variables
$f^{\gamma}_{\alpha\beta}=f^{\gamma}_{\alpha\beta}(p,q)$ and the
existence theorem for the nilpotent BRST charge has been proved
\cite{BFV}. It allows to present ${\cal Q}$ by series expansion
(in general, infinite) in ghost variables
\begin{eqnarray}\label{i2}
{\cal Q}=c^{\alpha}T_{\alpha}-\frac{1}{2}c^{\alpha}c^{\beta}
f^{\gamma}_{\alpha\beta}{\cal P}_{\gamma}+\cdot\cdot\cdot .
\end{eqnarray}
Here the dots mean the terms of higher orders in ghost variables
conditioned by $p,q$ dependence of structure functions.  The
problem, which we discuss here, consists in construction for a given
constrained theory the higher order contributions to ${\cal Q}$ in
terms of its structure functions. In general, solution to this
problem is unknown. We would like to point out specially that
although the existing theorem for BRST charge in general gauge
theories has been proved \cite{BFV} (see also the reviews
\cite{BF}), there is an independent problem of explicit finding
nilpotent BRST charge in closed form for concrete theory in terms of
its phase space dependent structure functions, i.e. a problem of
explicit solution of the general equations determining the BRST
charge in BFV approach \cite{BFV}.

Development of conformal field theories led to discovery of a new
class of gauge theories possessing the nonlinear gauge algebras,
called ${\cal W}_N$ algebras, where structure functions essentially
depend on the phase variables (see \cite{CFT} for ${\cal W}_3$
algebra and \cite{WN} for various generalizations). The BRST
construction for such algebras was discussed in \cite{tm},
\cite{SSvN}. Also we point out the closely related problem of
construction of the BRST charge for quantum groups with
quadratically non-linear algebras \cite{IO}. Recently, it was shown
that a special class of nonlinear gauge algebras arises in higher
spin theory on AdS space \cite{HSF} where the corresponding BRST
charge has also been found.

From a general point of view the nonlinear algebras considered in
\cite{CFT}, \cite{WN}, \cite{HSF} are characterized by the following
property. All of them can be described in terms of  constraints
$T_{\alpha}$ satisfying the relation (\ref{i0}) with nonconstant
structure functions which form a finite order polynomial in the
constraints $T_{\alpha}$
\begin{eqnarray}\label{j}
f^{\gamma}_{\alpha\beta}= F^{\;\;\gamma}_{\alpha\beta}+
V^{(1)\gamma\beta_1}_{\alpha\beta}T_{\beta_1}+\cdot\cdot\cdot+
V^{(n-1)\gamma\beta_1...\beta_{n-1}}_{\alpha\beta}T_{\beta_1}\cdot\cdot\cdot
T_{\beta_{n-1}}
\end{eqnarray}
where $F^{\;\;\gamma}_{\alpha\beta},
V^{(1)\gamma\beta_1}_{\alpha\beta},...,
V^{(n-1)\gamma\beta_1...\beta_{n-1}}_{\alpha\beta}$ are constants.
Construction of the nilpotent BRST charge for quadratically
nonlinear algebras ($ V^{(2)\gamma\beta_1\beta_2}_{\alpha\beta}=...=
V^{(n-1)\gamma\beta_1...\beta_{n-1}}_{\alpha\beta}=0$) subjected to
an additional special assumption concerning structure constants
$V^{(1)\gamma\delta}_{\alpha\beta}=V^{\gamma\delta}_{\alpha\beta}$
(see below) was performed in \cite{SSvN} with the result
\begin{eqnarray}\label{i3}
{\cal Q}=c^{\alpha}T_{\alpha}-\frac{1}{2}c^{\alpha}c^{\beta}
F^{\gamma}_{\alpha\beta}{\cal P}_{\gamma}-
\frac{1}{2}c^{\alpha}c^{\beta}V^{\gamma}_{\alpha\beta}{\cal
P}_{\gamma}- \frac{1}{24}c^{\alpha}c^{\beta}c^{\gamma}c^{\delta}
V^{\mu\nu}_{\alpha\beta}V^{\rho\sigma}_{\gamma\delta}F^{\lambda}_{\mu\rho}
{\cal P}_{\nu}{\cal P}_{\sigma}{\cal P}_{\lambda}.
\end{eqnarray}
As to general nonlinear algebras of the form (\ref{j}), to our
knowledge, the problem of construction of a nilpotent the BRST
charge is open in this case.

In the present paper, we investigate construction of the nilpotent
BRST charge for nonlinear algebras (\ref{j}) and find  some special
restrictions on structure constants when a nilpotent BRST charge can
be presented in the simplest form.

The paper is organized as follows. In Section 2 we construct the
classical nilpotent BRST charge for quadratically nonlinear algebras
of constraints. In Section 3 we extend our consideration of  a
classical nilpotent BRST charge  for arbitrary nonlinear algebras
and find the unique form of the BRST charge. In Section 4 concluding
remarks are given.

\section{BRST charge for quadratically nonlinear algebras}
Let us consider a set of constraints $T_{\alpha}=T_{\alpha}(p,q)$
which satisfy the following (quadratically nonlinear) algebra
\begin{eqnarray}\label{q1}
\{T_{\alpha},T_{\beta}\}=F^{\;\;\gamma}_{\alpha\beta}T_{\gamma}+
V^{\gamma\delta}_{\alpha\beta}T_{\delta}T_{\gamma}
\end{eqnarray}
where structure constants $F^{\;\;\gamma}_{\alpha\beta}$ and
$V^{\gamma\delta}_{\alpha\beta}$
obey the symmetry properties
\begin{eqnarray}\label{q2}
F^{\;\;\gamma}_{\alpha\beta}=-F^{\;\;\gamma}_{\beta\alpha},\quad
V^{\gamma\delta}_{\alpha\beta}=-V^{\gamma\delta}_{\beta\alpha}=
V^{\delta\gamma}_{\alpha\beta}.
\end{eqnarray}
The Jacobi identities for (\ref{q1}) read
\begin{eqnarray}\label{q3}
F^{\;\;\gamma}_{[\alpha\beta}F^{\;\;\delta}_{\lambda]\gamma}=0,\\
\label{q4}
F^{\;\;\rho}_{[\alpha\beta}V^{\mu\nu}_{\lambda]\rho}+
V^{\rho(\mu}_{[\alpha\beta}F^{\;\;\nu)}_{\lambda]\rho}=0,\\
\label{q5}
V^{\rho(\mu}_{[\alpha\beta}V^{\nu\gamma)}_{\lambda]\rho}=0,
\end{eqnarray}
where symbols $()$ and $[\;]$ denote symmetrization and antisymmetrization
with respect to indices included in these brackets respectively.

Construction of the classical BRST charge \cite{BV},\cite{BFV} for a
given set of constraints (\ref{q1}) involves introducing for each
constraint $T_{\alpha}$ an anticommuting ghost $c^{\alpha}$ and an
anticommuting momentum ${\cal P}_{\alpha}$ having the following
distribution of the Grassmann parity
$\varepsilon(c^{\alpha})=\varepsilon({\cal P}_{\alpha})=1$ and the
ghost number
 $gh(c^{\alpha})=-gh({\cal P}_{\alpha})=1$ and obeying the relations
\begin{eqnarray}\label{q6}
 \{c^{\alpha},{\cal P}_{\beta}\}=\delta^{\alpha}_{\beta},\quad
 \{c^{\alpha},c^{\beta}\}=0,\quad
 \{{\cal P}_{\alpha},{\cal P}_{\beta}\}=0,\quad
 \{c^{\alpha},T_{\beta}\}=0,\quad \{{\cal P}_{\alpha},T_{\beta}\}=0.
\end{eqnarray}
The classical BRST charge ${\cal Q}$ is defined as a solution to the equation
\begin{eqnarray}\label{Q}
\{{\cal Q},{\cal Q}\}=0
\end{eqnarray}
being odd function of variables $(p,q,c,{\cal P})$ with the ghost
number $gh({\cal Q})=1$ and satisfying the boundary condition
\begin{eqnarray}\label{q7}
\frac{\partial {\cal Q}}{\partial c^{\alpha}}\Big|_{c=0}=T_{\alpha}.
\end{eqnarray}
We look for a solution to this problem in the form of the
power-series expansions in the ghost variables
\begin{eqnarray}\label{q8}
{\cal
Q}=c^{\alpha}T_{\alpha}+\sum_{k=1}^{\infty}c^{\alpha_1}c^{\alpha_2}
\cdot\cdot\cdot c^{\alpha_{k+1}}
U^{(k)\beta_1\beta_2...\beta_k}_{\;\;\;\alpha_1\alpha_2...\alpha_{k+1}}
{\cal P}_{\beta_1}{\cal P}_{\beta_2}\cdot\cdot\cdot{\cal
P}_{\beta_k}=c^{\alpha}T_{\alpha}+\sum_{k=1}^{\infty}{\cal Q}_{k+1},
\end{eqnarray}
where the structure functions $U^{(k)}$ are totally antisymmetric in
both upper and lower indices.

In lower order we have the following requirement for ${\cal Q}$ to be
nilpotent
\begin{eqnarray}\label{q9}
c^{\alpha_1}c^{\alpha_2}\Big(\{T_{\alpha_1},T_{\alpha_2}\}+
2U^{(1)\beta_1}_{\alpha_1\alpha_2}
T_{\beta_1}\Big)=0\;\;\;\Rightarrow \;\;\;\{T_{\alpha_1},T_{\alpha_2}\}+
2U^{(1)\beta_1}_{\alpha_1\alpha_2}
T_{\beta_1}=0.
\end{eqnarray}
It leads to  the structure function $U^{(1)}$ in the form
\begin{eqnarray}\label{q10}
U^{(1)\gamma}_{\;\;\alpha\beta}=-\frac{1}{2}(F^{\;\;\gamma}_{\alpha\beta}+
V^{\gamma\delta}_{\alpha\beta}T_{\delta})
\end{eqnarray}
and the contribution of the second order in ghosts $c^{\alpha}$ for
${\cal Q}$
\begin{eqnarray}\label{Q2}
{\cal
Q}_2=-\frac{1}{2}c^{\alpha}c^{\beta}(F^{\;\;\gamma}_{\alpha\beta}+
V^{\gamma\delta}_{\alpha\beta}T_{\delta}){\cal P}_{\gamma}.
\end{eqnarray}
Condition of nilpotency for ${\cal Q}$ in the next order has the
form
\begin{eqnarray}\label{q11}
&&c^{\alpha_1}c^{\alpha_2}c^{\alpha_3}
\Big(\{U^{(1)\beta_1}_{\alpha_1\alpha_2},T_{\alpha_3}\}+
2U^{(1)\gamma}_{\alpha_1\alpha_2}U^{(1)\beta_1}_{\alpha_3\gamma}+
2U^{(2)\beta_2\beta_1}_{\;\;\alpha_1\alpha_2\alpha_3}T_{\beta_2}\Big)
{\cal P}_{\beta_1}=0.
\end{eqnarray}
Using (\ref{q1}) and Jacobi identities (\ref{q3})-(\ref{q5}) one can rewrite the
equation (\ref{q11}) as
\begin{eqnarray}\label{q12}
&&c^{\alpha_1}c^{\alpha_2}c^{\alpha_3}
\Big(-V^{\gamma\rho}_{\alpha_1\alpha_2}V^{\sigma\beta_1}_{\alpha_3\gamma}
T_{\rho}T_{\sigma}+
4U^{(2)\beta_2\beta_1}_{\;\;\alpha_1\alpha_2\alpha_3}T_{\beta_2}\Big)
{\cal P}_{\beta_1}=0
\end{eqnarray}
or in the form
\begin{eqnarray}\label{3q1}
&&c^{\alpha_1}c^{\alpha_2}c^{\alpha_3}
\Big(-V^{\gamma\rho}_{[\alpha_1\alpha_2}V^{\sigma\beta_1}_{\alpha_3]\gamma}
T_{\rho}T_{\sigma}+
12U^{(2)\beta_2\beta_1}_{\;\;\alpha_1\alpha_2\alpha_3}T_{\beta_2}\Big)
{\cal P}_{\beta_1}=0
\end{eqnarray}
Let us introduce the quantity
$K^{\alpha}_{\alpha_1\alpha_2\alpha_3}$
\begin{eqnarray}\label{q13}
K^{\alpha}_{\alpha_1\alpha_2\alpha_3}=
V^{\gamma\rho}_{[\alpha_1\alpha_2}V^{\sigma\alpha}_{\alpha_3]\gamma}
T_{\rho}T_{\sigma}.
\end{eqnarray}
From the Jacobi identities (\ref{q5}) it follows that
\begin{eqnarray}\label{q14}
K^{\alpha}_{\alpha_1\alpha_2\alpha_3}T_{\alpha}=0.
\end{eqnarray}
This means that $K^{\alpha}_{\alpha_1\alpha_2\alpha_3}$ can be
presented in the form
\begin{eqnarray}\label{q15}
K^{\alpha}_{\alpha_1\alpha_2\alpha_3}=K^{[\alpha\beta]}_{\alpha_1\alpha_2\alpha_3}
T_{\beta},\quad
K^{[\alpha\beta]}_{\alpha_1\alpha_2\alpha_3}=
-K^{[\beta\alpha]}_{\alpha_1\alpha_2\alpha_3}.
\end{eqnarray}
In its turn, $K^{[\alpha\beta]}_{\alpha_1\alpha_2\alpha_3}$ depends
on $T_{\alpha}$ linearly (see (\ref{q13}) and (\ref{q15}))
\begin{eqnarray}\label{q16}
K^{[\alpha\beta]}_{\alpha_1\alpha_2\alpha_3}=
K^{[\alpha\beta]\sigma}_{\alpha_1\alpha_2\alpha_3}
T_{\sigma}.
\end{eqnarray}
In terms of these quantities the structure functions $U^{(2)}$ read
\begin{eqnarray}\label{3q2}
U^{(2)\beta_1\beta_2}_{\;\;\alpha_1\alpha_2\alpha_3}=-\frac{1}{12}
K^{[\beta_1\beta_2]\sigma}_{\alpha_1\alpha_2\alpha_3} T_{\sigma}
\end{eqnarray}

 Taking into account  (\ref{q13}),
(\ref{q15}), (\ref{q16}) and the Jacobi identities (\ref{q5}) we
obtain the following equations to define the explicit form of
$K^{[\alpha\beta]\sigma}_{\alpha_1\alpha_2\alpha_3}$
\begin{eqnarray}\label{q17}
K^{[\alpha\beta]\sigma}_{\alpha_1\alpha_2\alpha_3}+
K^{[\alpha\sigma]\beta}_{\alpha_1\alpha_2\alpha_3}=
-V^{\gamma\alpha}_{[\alpha_1\alpha_2}V^{\beta\sigma}_{\alpha_3]\gamma}.
\end{eqnarray}
This is a basic equation for determining the quantity $U^{(2)}$
(\ref{3q2}).

To solve the equations (\ref{q17}) we note that if it has a solution
$ K^{[\alpha\beta]\sigma}_{\alpha_1\alpha_2\alpha_3}$ then
\begin{eqnarray}\label{x1}
K^{[\alpha\beta]\sigma}_{\alpha_1\alpha_2\alpha_3}+
X^{[\alpha\beta\sigma]}_{[\alpha_1\alpha_2\alpha_3]}
\end{eqnarray}
will be a solution to (\ref{q17}) as well. In (\ref{x1})
$X^{[\alpha\beta\sigma]}_{[\alpha_1\alpha_2\alpha_3]}$ is totally
antisymmetric in both upper and lower indices. It is known \cite{BF}
that this arbitrariness can be removed by a canonical transformation
of the BRST charge which does not change the boundary condition
(\ref{q7}). From (\ref{q13}) and (\ref{q17}) one can try to find
$K^{[\alpha\beta]\sigma}_{\alpha_1\alpha_2\alpha_3}$ as a linear
combination of structures
$V^{\gamma\alpha}_{[\alpha_1\alpha_2}V^{\beta\sigma}_{\alpha_3]\gamma}$
(in Appendix A we explicitly verify this proposal as well as the
arbitrariness in solutions for the simplest case of three
constraints in the algebra (\ref{q1})). Therefore we can propose
\begin{eqnarray}\label{q18}
K^{[\alpha\beta]\sigma}_{\alpha_1\alpha_2\alpha_3}=
C^{[\alpha\beta]\sigma}_{\lambda(\mu\nu)}
V^{\gamma\lambda}_{[\alpha_1\alpha_2}V^{\mu\nu}_{\alpha_3]\gamma}
\end{eqnarray}
where $C^{[\alpha\beta]\sigma}_{\lambda(\mu\nu)}$ is a matrix constructed from
the unit matrices $\delta^{\alpha}_{\mu}$. It is not difficult to find
a general structure of $C^{[\alpha\beta]\sigma}_{\lambda(\mu\nu)}$ with
the required symmetry properties
\begin{eqnarray}\label{q19}
C^{[\alpha\beta]\sigma}_{\lambda(\mu\nu)}=
C\Big(\delta^{\alpha}_{\lambda}\delta^{\beta}_{\mu}\delta^{\sigma}_{\nu}+
\delta^{\alpha}_{\lambda}\delta^{\sigma}_{\mu}\delta^{\beta}_{\nu}-
\delta^{\alpha}_{\mu}\delta^{\beta}_{\lambda}\delta^{\sigma}_{\nu}-
\delta^{\alpha}_{\nu}\delta^{\beta}_{\lambda}\delta^{\sigma}_{\mu} \Big)
\end{eqnarray}
where $C$ is a constant. Returning with this result in (\ref{q17}) and using
the Jacobi identities (\ref{q5}) we obtain
\begin{eqnarray}\label{q19}
(6C+1)V^{\gamma\lambda}_{[\alpha_1\alpha_2}V^{\mu\nu}_{\alpha_3]\gamma}=0.
\end{eqnarray}
Therefore to this order we have two solutions to the nilpotency
condition. The first one corresponds to
\begin{eqnarray}\label{3sq1}
C=-\frac{1}{6}
\end{eqnarray}
and leads to
\begin{eqnarray}\label{q20}
K^{[\alpha\beta]\sigma}_{\alpha_1\alpha_2\alpha_3}=
-\frac{1}{3}\Big(
V^{\gamma\alpha}_{[\alpha_1\alpha_2}V^{\beta\sigma}_{\alpha_3]\gamma}-
V^{\gamma\beta}_{[\alpha_1\alpha_2}V^{\alpha\sigma}_{\alpha_3]\gamma}\Big)=
-\frac{1}{3}V^{\gamma[\alpha}_{[\alpha_1\alpha_2}
V^{\beta]\sigma}_{\alpha_3]\gamma}.
\end{eqnarray}
Therefore
\begin{eqnarray}\label{q21}
U^{(2)\beta_1\beta_2}_{\alpha_1\alpha_2\alpha_3}=\frac{1}{36}
V^{\gamma[\beta_1}_{[\alpha_1\alpha_2}V^{\beta_2]\sigma}_{\alpha_3]\gamma}
T_{\sigma}
\end{eqnarray}
and
\begin{eqnarray}\label{3Q}
{\cal Q}_3=\frac{1}{6}c^{\alpha_1}c^{\alpha_2}c^{\alpha_3}
V^{\gamma\beta_1}_{\alpha_1\alpha_2}V^{\beta_2\sigma}_{\alpha_3\gamma}
T_{\sigma}{\cal P}_{\beta_1}{\cal P}_{\beta_2}
\end{eqnarray}

 The second possibility corresponds to restriction on structure
functions of nonlinear algebras
\begin{eqnarray}\label{R1}
V^{\gamma\lambda}_{[\alpha_1\alpha_2}V^{\mu\nu}_{\alpha_3]\gamma}=0.
\end{eqnarray}
This means that
$K^{[\alpha\beta]\sigma}_{\alpha_1\alpha_2\alpha_3}=0$ and
\begin{eqnarray}\label{q23}
U^{(2)\beta_1\beta_2}_{\alpha_1\alpha_2\alpha_3}=0,\quad {\cal
Q}_3=0.
\end{eqnarray}
In  \cite{SSvN} a class of quadratically nonlinear algebras
subjected to restrictions
\begin{eqnarray}\label{vNc}
V^{\gamma\lambda}_{\alpha_1\alpha_2}V^{\mu\nu}_{\alpha_3\gamma}=0.
\end{eqnarray}
was investigated to construct a nilpotent BRST charge. These
restrictions look more stronger in comparison with (\ref{R1}). They
are not the direct consequence of (\ref{R1}) and look like
additional restrictions which are not dictated by solutions to the
nilpotency conditions in the third order. We point out that the
Jacobi identities (\ref{q5}) are satisfied in the case of conditions
(\ref{R1}). In what follows we suppose fulfilment of conditions
(\ref{R1}).

Using these results we rewrite the condition of nilpotency in the
forth order in the ghost fields $c$ as follows
\begin{eqnarray}
\label{q16}
c^{\alpha_1}c^{\alpha_2}c^{\alpha_3}c^{\alpha_4}
\Big(\{U^{(1)\beta_1}_{\alpha_1\alpha_2},U^{(1)\beta_2}_{\alpha_3\alpha_4}\}+
6U^{(3)\beta_1\beta_2\beta_3}_{\;\;\alpha_1\alpha_2\alpha_3\alpha_4}
T_{\beta_3}\Big){\cal P}_{\beta_1}{\cal P}_{\beta_2}=0,
\end{eqnarray}
or
\begin{eqnarray}\label{c4}
c^{\alpha_1}c^{\alpha_2}c^{\alpha_3}c^{\alpha_4}
\Big(V^{\beta\beta_1}_{\alpha_1\alpha_2}
V^{\beta_2\gamma}_{\alpha_3\alpha_4}F^{\;\;\beta_3}_{\beta\gamma}+
V^{\beta\beta_1}_{\alpha_1\alpha_2}
V^{\beta_2\gamma}_{\alpha_3\alpha_4}
V^{\beta_3\sigma}_{\beta\gamma}T_{\sigma}+
24U^{(3)\beta_1\beta_2\beta_3}_{\;\;\alpha_1\alpha_2\alpha_3\alpha_4}\big)
T_{\beta_3}{\cal P}_{\beta_1}{\cal P}_{\beta_2}=0.
\end{eqnarray}

To define the structure function $U^{(3)}$ correctly we have to antisymmetrize
quantities presented in (\ref{c4}) in indices
$\alpha_1,\alpha_2,\alpha_3,\alpha_4$ and $\beta_1,\beta_2,\beta_3$. Taking into
account that expressions in (\ref{q16}) are antisymmetrical in
$\beta_1,\beta_2$ the antisymmetrization in $\beta_1,\beta_2,\beta_3$ is trivial
\begin{eqnarray}\nonumber
[123]=[[12]3]]=[12]3+[23]1+[31]2=[12]3+cycle(123).
\end{eqnarray}
The antisymmetrization in $\alpha_1,\alpha_2,\alpha_3,\alpha_4$ is
performed for two sets of antisymmetric indices $[\alpha_1\alpha_2]$
and $[\alpha_3\alpha_4]$. Symbolically, we can write this procedure
as follows
\begin{eqnarray}\nonumber
[1234]=[[12][34]]=[12][34]+[23][14]+[31][24]+[24][31]+[34][12]+[14][23].
\end{eqnarray}
Then from (\ref{c4}) we obtain the equations for structure functions
 $U^{(3)}$
\begin{eqnarray}
V^{\beta\beta_1}_{[\alpha_1\alpha_2}
V^{\beta_2\gamma}_{\alpha_3\alpha_4]}F^{\;\;\beta_3}_{\beta\gamma}T_{\beta_3}+
V^{\beta\beta_1}_{[\alpha_1\alpha_2}
V^{\beta_2\gamma}_{\alpha_3\alpha_4]}
V^{\beta_3\sigma}_{\beta\gamma}T_{\sigma}T_{\beta_3}+
144U^{(3)\beta_1\beta_2\beta_3}_{\;\;\alpha_1\alpha_2\alpha_3\alpha_4}
T_{\beta_3}=0,
\end{eqnarray}
where the following symmetry properties
\begin{eqnarray}\label{3rdR1}
V^{\beta\beta_1}_{[\alpha_1\alpha_2}
V^{\beta_2\gamma}_{\alpha_3\alpha_4]}F^{\;\;\beta_3}_{\beta\gamma}=-
V^{\beta\beta_2}_{[\alpha_1\alpha_2}
V^{\beta_1\gamma}_{\alpha_3\alpha_4]}F^{\;\;\beta_3}_{\beta\gamma},\quad
V^{\beta\beta_1}_{[\alpha_1\alpha_2}
V^{\beta_2\gamma}_{\alpha_3\alpha_4]}
V^{\beta_3\sigma}_{\beta\gamma}=-
V^{\beta\beta_2}_{[\alpha_1\alpha_2}
V^{\beta_1\gamma}_{\alpha_3\alpha_4]}
V^{\beta_3\sigma}_{\beta\gamma},
\end{eqnarray}
were used. Moreover, the following conditions of consistency are
satisfied
\begin{eqnarray}\label{3rdR2}
V^{\beta\beta_1}_{[\alpha_1\alpha_2}
V^{\gamma(\beta_2}_{\alpha_3\alpha_4]}F^{\;\;\beta_3)}_{\beta\gamma}=0,\quad
V^{\beta\beta_1}_{[\alpha_1\alpha_2}
V^{\gamma(\beta_2}_{\alpha_3\alpha_4]}
V^{\beta_3\sigma)}_{\beta\gamma}=0
\end{eqnarray}
which are consequence of the Jacobi identities and consistent with
 (\ref{R1}) (see Appendix B). From (\ref{3rdR1}) and
(\ref{3rdR2}) it follows that $V^{\beta\beta_1}_{[\alpha_1\alpha_2}
V^{\beta_2\gamma}_{\alpha_3\alpha_4]}F^{\;\;\beta_3}_{\beta\gamma}$
are totally antisymmetric in indices $\beta_1,\beta_2,\beta_3,$ and
repeat the symmetry properties of
$U^{(3)\beta_1\beta_2\beta_3}_{\;\;\alpha_1\alpha_2\alpha_3\alpha_4}$.
Moreover, when restrictions (\ref{R1}) are satisfied then one can
prove (see Appendix B) that
\begin{eqnarray}\label{Z}
V^{\beta\beta_1}_{[\alpha_1\alpha_2}
V^{\gamma\beta_2}_{\alpha_3\alpha_4]}
V^{\beta_3\sigma}_{\beta\gamma}=0.
\end{eqnarray}
Then, finally we get the solution for the structure functions
$U^{(3)}$
\begin{eqnarray}
U^{(3)\beta_1\beta_2\beta_3}_{\;\;\alpha_1\alpha_2\alpha_3\alpha_4}=
-\frac{1}{144}V^{\beta\beta_1}_{[\alpha_1\alpha_2}
V^{\beta_2\gamma}_{\alpha_3\alpha_4]}F^{\;\;\beta_3}_{\beta\gamma}
\end{eqnarray}
and for the BRST charge in the forth order
\begin{eqnarray}
{\cal Q}_4=-\frac{1}{24}c^{\alpha_1}c^{\alpha_2}c^{\alpha_3}c^{\alpha_4}
V^{\beta\beta_1}_{\alpha_1\alpha_2}
V^{\beta_2\gamma}_{\alpha_3\alpha_4}
F^{\;\;\beta_3}_{\beta\gamma}{\cal P}_{\beta_1}
{\cal P}_{\beta_2}{\cal P}_{\beta_3}.
\end{eqnarray}

Therefore for any theory with quadratically nonlinear algebra
(\ref{q1}) subjected to the Jacobi identities (\ref{q3}) and
(\ref{q4}) as well as the additional restrictions (\ref{R1}) and
\begin{eqnarray}\label{qq}
V^{\beta\beta_1}_{[\alpha_1\alpha_2}
V^{\beta_2\gamma}_{\alpha_3\alpha_4]}F^{\;\;\beta_3}_{\beta\gamma}=0
\end{eqnarray}
 there
exists a unique form of nilpotent the BRST charge
\begin{eqnarray}\label{q24}
{\cal Q}=c^{\alpha}T_{\alpha}-\frac{1}{2}c^{\alpha}c^{\beta}
(F^{\;\;\gamma}_{\alpha\beta}+
V^{\gamma\delta}_{\alpha\beta}T_{\delta}){\cal P}_{\gamma}.
\end{eqnarray}

\section{BRST charge for generic nonlinear algebras}

Nonlinear algebras are defined by the following relations
\begin{eqnarray}\label{n1}
\{T_{\alpha},T_{\beta}\}=F^{\;\;\gamma}_{\alpha\beta}T_{\gamma}+
V^{(1)\alpha_1\alpha_2}_{\alpha\beta}T_{\alpha_1}T_{\alpha_2}+
V^{(2)\alpha_1\alpha_2\alpha_3}_{\alpha\beta}T_{\alpha_1}T_{\alpha_2}T_{\alpha_3}
+\cdot\cdot\cdot
+V^{(n-1)\alpha_1...\alpha_n}_{\alpha\beta}T_{\alpha_1}\cdot\cdot\cdot
T_{\alpha_n},
\end{eqnarray}
where structure constants $F^{\;\;\gamma}_{\alpha\beta}$ and
$V^{(k-1)\alpha_1...\alpha_k}_{\alpha\beta}\;(k=2,3,...,n)$ are
antisymmetric in lower indices and
$V^{(k-1)\alpha_1...\alpha_k}_{\alpha\beta}\;(k=2,3,...,n)$ are
totally symmetric in upper indices.

The Jacobi identities for (\ref{n1}) have the form
\begin{eqnarray}\label{n2}
&&F^{\;\;\gamma}_{[\alpha\beta}F^{\;\;\delta}_{\lambda]\gamma}=0\;,\\
\label{n}
&&F^{\;\;\rho}_{[\alpha\beta}V^{(1)\beta_1\beta_2}_{\lambda]\rho} +
V^{(1)\rho(\beta_1}_{[\alpha\beta} F^{\;\;\beta_2)}_{\lambda]\rho}=0
\;,\\
\nonumber
&&F^{\;\;\rho}_{[\alpha\beta}V^{(m)\beta_1...\beta_m\beta_{m+1}}_{\lambda]\rho}
+
V^{(m)\rho(\beta_1...\beta_m}_{[\alpha\beta}
F^{\;\;\beta_{m+1})}_{\lambda]\rho}+\\
\label{n3}
&&+\sum_{k=1}^{m-1}\frac{(k+1)!(m-k+1)!}{(m+1)!}
V^{(k)\rho(\beta_1...\beta_k}_{[\alpha\beta}
V^{(m-k)\beta_{k+1}...\beta_{m+1})}_{\lambda]\rho}
=0\; (m=2,3,...,n-1)\;, \\
\label{n4}
&&\sum_{k=m-n+1}^{n-1}\frac{(k+1)!(m-k+1)!}{(m+1)!}
V^{(k)\rho(\beta_1...\beta_{k}}_{[\alpha\beta}
V^{(m-k)\beta_{k+1}...\beta_{m+1})}_{\lambda]\rho}=0\;(m=n,...,2n-2)\;.
\end{eqnarray}
In Eqs. (\ref{n3}), (\ref{n4}) symmetrization includes  two sets of
symmetric indices. We assume that in the symmetrization only one
representative among equivalent ones obtained by permutation of
indices into these sets is presented.

Consider now construction of the classical BRST charge ${\cal Q}$
for the algebra (\ref{n1}). Equation for the structure function $U^{(1)}$
has the form
\begin{eqnarray}\label{n5}
c^{\alpha}c^{\beta}\Big(\{T_{\alpha},T_{\beta}\}+
2U^{(1)\gamma}_{\alpha\beta}T_{\gamma}\Big)=0
\end{eqnarray}
or equivalently
\begin{eqnarray}\label{n6}
c^{\alpha}c^{\beta}\Big(F^{\;\;\gamma}_{\alpha\beta}+
V^{(1)\gamma\alpha_1}_{\alpha\beta}T_{\alpha_1}+\cdot\cdot\cdot +
V^{(n-1)\gamma\alpha_1...\alpha_{n-1}}_{\alpha\beta}T_{\alpha_1}\cdot\cdot\cdot
T_{\alpha_{n-1}}
+2U^{(1)\gamma}_{\;\;\alpha\beta}\Big)T_{\gamma}=0
\end{eqnarray}
with the evident solution
\begin{eqnarray}\label{n7}
U^{(1)\gamma}_{\;\;\alpha\beta}=-\frac{1}{2}\Big(F^{\;\;\gamma}_{\alpha\beta}+
V^{(1)\gamma\alpha_1}_{\alpha\beta}T_{\alpha_1}+\cdot\cdot\cdot+
V^{(n-1)\gamma\alpha_1...\alpha_{n-1}}_{\alpha\beta}T_{\alpha_1}\cdot\cdot\cdot
T_{\alpha_{n-1}}\Big).
\end{eqnarray}
In the next order we have the following equation of nilpotency
\begin{eqnarray}\label{n8}
c^{\alpha_1}c^{\alpha_2}c^{\alpha_3}\Big(
\{U^{(1)\beta_1}_{\alpha_1\alpha_2},T_{\alpha_3}\}+
2U^{(1)\gamma}_{\alpha_1\alpha_2}
U^{(1)\beta_1}_{\alpha_3\gamma}+
2U^{(2)\gamma\beta_1}_{\alpha_1\alpha_2\alpha_3}T_{\gamma} \Big)
{\cal P}_{\beta_1}=0.
\end{eqnarray}
To solve this equation for the structure function $U^{(2)}$ let us consider
the equality
\begin{eqnarray}\label{n9}
\{T_{\alpha_1},T_{\alpha_2}\}+
2U^{(1)\beta_1}_{\alpha_1\alpha_2}T_{\beta_1}=0
\end{eqnarray}
and the consequence from it
\begin{eqnarray}\label{n10}
\{\{T_{\alpha_1},T_{\alpha_2}\},T_{\alpha_3}\}+
2U^{(1)\beta_1}_{\alpha_1\alpha_2}\{T_{\beta_1},T_{\alpha_3}\}+
2\{U^{(1)\beta_1}_{\alpha_1\alpha_2},T_{\alpha_3}\}T_{\beta_1}=0.
\end{eqnarray}
Using Eq.(\ref{n9}) and the symmetry properties of $U^{(1)}$ we can
rewrite the last equation asfollows
\begin{eqnarray}\label{n11}
\{\{T_{\alpha_1},T_{\alpha_2}\},T_{\alpha_3}\}+
2\Big(2U^{(1)\gamma}_{\alpha_1\alpha_2}U^{(1)\beta_1}_{\alpha_3\gamma}+
\{U^{(1)\beta_1}_{\alpha_1\alpha_2},T_{\alpha_3}\}\Big) T_{\beta_1}=0.
\end{eqnarray}
Taking into account the Jacobi identities for Poisson bracket (see
(\ref{n2})--(\ref{n4}))
$\{\{T_{\alpha_1},T_{\alpha_2}\},T_{\alpha_3}\}+cycle(\alpha_1\alpha_2\alpha_3)=0
$, from (\ref{n11}) it follows that
\begin{eqnarray}\label{n12}
\Big(2U^{(1)\gamma}_{[\alpha_1\alpha_2}U^{(1)\beta_1}_{\alpha_3]\gamma}+
\{U^{(1)\beta_1}_{[\alpha_1\alpha_2},T_{\alpha_3]}\}\Big)T_{\beta_1}=0.
\end{eqnarray}
Let us introduce the quantities
\begin{eqnarray}\label{m1}
K^{\beta_1}_{\alpha_1\alpha_2\alpha_3}=
2U^{(1)\gamma}_{[\alpha_1\alpha_2}U^{(1)\beta_1}_{\alpha_3]\gamma}+
\{U^{(1)\beta_1}_{[\alpha_1\alpha_2},T_{\alpha_3]}\}
\end{eqnarray}
obeying the property
\begin{eqnarray}\label{m2}
K^{\beta_1}_{\alpha_1\alpha_2\alpha_3}T_{\beta_1}=0
\end{eqnarray}
due to (\ref{n12}). This means that
$K^{\beta_1}_{\alpha_1\alpha_2\alpha_3}$ can be presented in the
form
\begin{eqnarray}\label{m3}
K^{\beta_1}_{\alpha_1\alpha_2\alpha_3}=
K^{[\beta_1\beta_2]}_{\alpha_1\alpha_2\alpha_3}T_{\beta_2},\quad
K^{[\beta_1\beta_2]}_{\alpha_1\alpha_2\alpha_3}=-
K^{[\beta_2\beta_1]}_{\alpha_1\alpha_2\alpha_3}.
\end{eqnarray}
In terms of $K^{[\beta_1\beta_2]}_{\alpha_1\alpha_2\alpha_3}$ solution for
structure functions $U^{(2)\beta_1\beta_2}_{\alpha_1\alpha_2\alpha_3}$ reads
\begin{eqnarray}\label{m4}
U^{(2)\beta_1\beta_2}_{\alpha_1\alpha_2\alpha_3}=\frac{1}{6}
K^{[\beta_1\beta_2]}_{\alpha_1\alpha_2\alpha_3}.
\end{eqnarray}
Straightforward calculations of
$K^{\beta_1}_{\alpha_1\alpha_2\alpha_3}$ lead to the following
result
\begin{eqnarray}\label{m5}\nonumber
K^{\beta_1}_{\alpha_1\alpha_2\alpha_3}&=&\frac{1}{2}
\Big(F^{\sigma}_{[\alpha_1\alpha_2}F^{\beta_1}_{\alpha_3]\sigma}+
\big(F^{\sigma}_{[\alpha_1\alpha_2}V^{(1)\beta_1\sigma_1}_{\alpha_3]\sigma}
+V^{(1)\sigma(\sigma_1}_{[\alpha_1\alpha_2}
F^{\beta_1)}_{\alpha_3]\sigma}\big)T_{\sigma_1}+\\
\nonumber
&+&\sum_{m=2}^{n-1}\big(F^{\sigma}_{[\alpha_1\alpha_2}
V^{(m)\beta_1\sigma_1...\sigma_m}_{\alpha_3]\sigma}
+V^{(m)\sigma(\sigma_1...\sigma_m}_{[\alpha_1\alpha_2}
F^{\beta_1)}_{\alpha_3]\sigma}\big)T_{\sigma_1}\cdot\cdot\cdot T_{\sigma_m}+\\
\nonumber
&+&\sum_{m=2}^{n-1}\sum_{k=1}^{m-1}\frac{k!(m-k)!}{m!}
V^{(k)\sigma(\beta_1\sigma_1...\sigma_{k-1}}_{[\alpha_1\alpha_2}
V^{(m-k)\sigma_{k}...\sigma_m)}_{\alpha_3]\sigma}\;T_{\sigma_1}
\cdot\cdot\cdot T_{\sigma_m}+\\
\nonumber
&+&\sum_{m=2}^{n-1}\sum_{k=1}^{m-1}\frac{k!(m-k)!(m-k)}{m!}
V^{(k)\sigma\beta_1(\sigma_1...\sigma_{k-1}}_{[\alpha_1\alpha_2}
V^{(m-k)\sigma_{k}...\sigma_m)}_{\alpha_3]\sigma}\;T_{\sigma_1}
\cdot\cdot\cdot T_{\sigma_m}+\\
\nonumber
&+&\sum_{m=n+1}^{2n-2}\sum_{k=m-n+1}^{n-1}\frac{k!(m-k)!}{m!}
V^{(k)\sigma(\beta_1\sigma_1...\sigma_{k-1}}_{[\alpha_1\alpha_2}
V^{(m-k)\sigma_{k}...\sigma_m)}_{\alpha_3]\sigma}\;T_{\sigma_1}
\cdot\cdot\cdot T_{\sigma_m}+\\
&+&\sum_{m=n+1}^{2n-2}\sum_{k=m-n+1}^{n-1}\frac{k!(m-k)!(m-k)}{m!}
V^{(k)\sigma\beta_1(\sigma_1...\sigma_{k-1}}_{[\alpha_1\alpha_2}
V^{(m-k)\sigma_{k}...\sigma_m)}_{\alpha_3]\sigma}\;T_{\sigma_1}
\cdot\cdot\cdot T_{\sigma_m}
\Big)\;.
\end{eqnarray}
In deriving (\ref{m5})  the relations
\begin{eqnarray}\label{m6}
&&V^{(k)\sigma\beta_1(\sigma_1...\sigma_{k}}_{[\alpha_1\alpha_2}
V^{(m-k)\sigma_{k+1}...\sigma_m)\beta_1}_{\alpha_3]\sigma}+
(m-k+1)V^{(k)\sigma\beta_1(\sigma_1...\sigma_{k-1}}_{[\alpha_1\alpha_2}
V^{(m-k)\sigma_{k}...\sigma_m)}_{\alpha_3]\sigma}=\\
\nonumber
&&=V^{(k)\sigma(\beta_1\sigma_1...\sigma_{k-1}}_{[\alpha_1\alpha_2}
V^{(m-k)\sigma_{k}...\sigma_m)}_{\alpha_3]\sigma}+
(m-k)V^{(k)\sigma\beta_1(\sigma_1...\sigma_{k-1}}_{[\alpha_1\alpha_2}
V^{(m-k)\sigma_{k}...\sigma_m)}_{\alpha_3]\sigma}
\end{eqnarray}
were used. Taking into account the Jacobi identities (\ref{n2})--
(\ref{n3}) we can rewrite (\ref{m5}) in the form
\begin{eqnarray}\label{m7}\nonumber
K^{\beta_1}_{\alpha_1\alpha_2\alpha_3}&=&\frac{1}{2} \Big(-
\sum_{m=2}^{n-1}\sum_{k=1}^{m-1}\frac{k!k\;(m-k)!(m-k)}{(m+1)!}
V^{(k)\sigma(\beta_1\sigma_1...\sigma_{k-1}}_{[\alpha_1\alpha_2}
V^{(m-k)\sigma_{k}...\sigma_m)}_{\alpha_3]\sigma}\;T_{\sigma_1}
\cdot\cdot\cdot T_{\sigma_m}+\\
\nonumber
&+&\sum_{m=2}^{n-1}\sum_{k=1}^{m-1}\frac{k!(m-k)!(m-k)}{m!}
V^{(k)\sigma\beta_1(\sigma_1...\sigma_{k-1}}_{[\alpha_1\alpha_2}
V^{(m-k)\sigma_{k}...\sigma_m)}_{\alpha_3]\sigma}\;T_{\sigma_1}
\cdot\cdot\cdot T_{\sigma_m}+\\
\nonumber
&+&\sum_{m=n+1}^{2n-2}\sum_{k=m-n+1}^{n-1}\frac{k!(m-k)!}{m!}
V^{(k)\sigma(\beta_1\sigma_1...\sigma_{k-1}}_{[\alpha_1\alpha_2}
V^{(m-k)\sigma_{k}...\sigma_m)}_{\alpha_3]\sigma}\;T_{\sigma_1}
\cdot\cdot\cdot T_{\sigma_m}+\\
&+&\sum_{m=n+1}^{2n-2}\sum_{k=m-n+1}^{n-1}\frac{k!(m-k)!(m-k)}{m!}
V^{(k)\sigma\beta_1(\sigma_1...\sigma_{k-1}}_{[\alpha_1\alpha_2}
V^{(m-k)\sigma_{k}...\sigma_m)}_{\alpha_3]\sigma}\;T_{\sigma_1}
\cdot\cdot\cdot T_{\sigma_m} \Big)\;.
\end{eqnarray}

Now let us  require the following additional restrictions on structure constants
of the algebra (\ref{n1})
\begin{eqnarray}\label{m8}
&&V^{(k)\sigma\beta_1\sigma_1...\sigma_{k-1}}_{[\alpha_1\alpha_2}
V^{(m-k)\sigma_{k}...\sigma_m}_{\alpha_3]\sigma}=0,\\
\nonumber
&&k=1,2,...,n-1,\;\; m>k,\;\; m=2,3,...,2n-2.
\end{eqnarray}
Then we obtain
\begin{eqnarray}\label{m9}
K^{\beta_1}_{\alpha_1\alpha_2\alpha_3}=0,\quad
U^{(2)\beta_1\beta_2}_{\alpha_1\alpha_2\alpha_3}=0,\quad {\cal
Q}_{3}=0.
\end{eqnarray}
Conditions (\ref{m8}) generalize (\ref{R1}) for nonlinear algebras.

With this result we have the following condition of nilpotency in
the next order
\begin{eqnarray}\label{n16}\nonumber
c^{\alpha_1}c^{\alpha_2}c^{\alpha_3}c^{\alpha_4}
&&\Big(\bar{V}^{\sigma\beta_1}_{\alpha_1\alpha_2}
\bar{V}^{\beta_2\rho}_{\alpha_3\alpha_4}F^{\beta_3}_{\sigma\rho}+
\bar{V}^{\sigma\beta_1}_{\alpha_1\alpha_2}
\bar{V}^{\beta_2\rho}_{\alpha_3\alpha_4}
\bar{V}^{\beta_3\lambda}_{\sigma\rho}T_{\lambda}+
\bar{V}^{\sigma[\beta_1}_{\alpha_1\alpha_2}
\tilde{V}^{\beta_2]\rho}_{\alpha_3\alpha_4}F^{\beta_3}_{\sigma\rho}+
\bar{V}^{\sigma[\beta_1}_{\alpha_1\alpha_2}
\tilde{V}^{\beta_2]\rho}_{\alpha_3\alpha_4}
\bar{V}^{\beta_3\lambda}_{\sigma\rho}T_{\lambda}+\\
&&+\tilde{V}^{\sigma\beta_1}_{\alpha_1\alpha_2}
\tilde{V}^{\beta_2\rho}_{\alpha_3\alpha_4}F^{\beta_3}_{\sigma\rho}+
\tilde{V}^{\sigma\beta_1}_{\alpha_1\alpha_2}
\tilde{V}^{\beta_2\rho}_{\alpha_3\alpha_4}
\bar{V}^{\beta_3\lambda}_{\sigma\rho}T_{\lambda}+
24U^{(3)\beta_1\beta_2\beta_3}_{\;\;\alpha_1\alpha_2\alpha_3\alpha_4}\big)
T_{\beta_3}{\cal P}_{\beta_1}{\cal P}_{\beta_2}=0
\end{eqnarray}
where the notations
\begin{eqnarray}
\label{n18}
\bar{V}^{\alpha\beta}_{\mu\nu}&=&\sum_{k=1}^{n-1}
V^{(k)\alpha\beta\sigma_1...\sigma_{k-1}}_{\mu\nu}T_{\sigma_1}
\cdot\cdot\cdot T_{\sigma_{k-1}}\\
\label{n17}
\tilde{V}^{\alpha\beta}_{\mu\nu}&=&\sum_{k=1}^{n-1}
(k-1)V^{(k)\alpha\beta\sigma_1...\sigma_{k-1}}_{\mu\nu}T_{\sigma_1}
\cdot\cdot\cdot T_{\sigma_{k-1}}.
\end{eqnarray}
were used. From (\ref{n16}) we obtain the following equations to
find
$U^{(3)\beta_1\beta_2\beta_3}_{\;\;\alpha_1\alpha_2\alpha_3\alpha_4}$
\begin{eqnarray}
\label{n19}\nonumber
\bar{V}^{\sigma\beta_1}_{[\alpha_1\alpha_2}
\bar{V}^{\beta_2\rho}_{\alpha_3\alpha_4]}F^{\beta_3}_{\sigma\rho}T_{\beta_3}+
\bar{V}^{\sigma\beta_1}_{[\alpha_1\alpha_2}
\bar{V}^{\beta_2\rho}_{\alpha_3\alpha_4]}
\bar{V}^{\beta_3\lambda}_{\sigma\rho}T_{\lambda}T_{\beta_3}+
\bar{V}^{\sigma[\beta_1}_{[\alpha_1\alpha_2}
\tilde{V}^{\beta_2]\rho}_{\alpha_3\alpha_4]}F^{\beta_3}_{\sigma\rho}T_{\beta_3}+
\bar{V}^{\sigma[\beta_1}_{[\alpha_1\alpha_2}
\tilde{V}^{\beta_2]\rho}_{\alpha_3\alpha_4]}
\bar{V}^{\beta_3\lambda}_{\sigma\rho}T_{\lambda}T_{\beta_3}+\\
+\tilde{V}^{\sigma\beta_1}_{[\alpha_1\alpha_2}
\tilde{V}^{\beta_2\rho}_{\alpha_3\alpha_4]}F^{\beta_3}_{\sigma\rho}T_{\beta_3}+
\tilde{V}^{\sigma\beta_1}_{[\alpha_1\alpha_2}
\tilde{V}^{\beta_2\rho}_{\alpha_3\alpha_4]}
\bar{V}^{\beta_3\lambda}_{\sigma\rho}T_{\lambda}T_{\beta_3}+
144U^{(3)\beta_1\beta_2\beta_3}_{\;\;\alpha_1\alpha_2\alpha_3\alpha_4}
T_{\beta_3}=0.
\end{eqnarray}
The conditions of consistency have the form
\begin{eqnarray}
\label{n20}
\bar{V}^{\sigma\beta_1}_{[\alpha_1\alpha_2}
\bar{V}^{\rho(\beta_2}_{\alpha_3\alpha_4]}F^{\beta_3)}_{\sigma\rho}=0,\quad
\tilde{V}^{\sigma\beta_1}_{[\alpha_1\alpha_2}
\tilde{V}^{\rho(\beta_2}_{\alpha_3\alpha_4]}F^{\beta_3)}_{\sigma\rho}=0,\quad
\bar{V}^{\sigma[\beta_1}_{[\alpha_1\alpha_2}
\tilde{V}^{\beta_2]\rho}_{\alpha_3\alpha_4]}F^{\beta_3}_{\sigma\rho}+
\bar{V}^{\sigma[\beta_1}_{[\alpha_1\alpha_2}
\tilde{V}^{\beta_3]\rho}_{\alpha_3\alpha_4]}F^{\beta_2}_{\sigma\rho}=0,\\
\label{n21}
\bar{V}^{\sigma\beta_1}_{[\alpha_1\alpha_2}
\bar{V}^{\rho(\beta_2}_{\alpha_3\alpha_4]}
\bar{V}^{\beta_3\lambda)}_{\sigma\rho}=0,\quad
\tilde{V}^{\sigma\beta_1}_{[\alpha_1\alpha_2}
\tilde{V}^{\rho(\beta_2}_{\alpha_3\alpha_4]}
\bar{V}^{\beta_3\lambda)}_{\sigma\rho}=0,\quad
\bar{V}^{\sigma[\beta_1}_{[\alpha_1\alpha_2}
\tilde{V}^{\beta_2]\rho}_{\alpha_3\alpha_4]}
\bar{V}^{\beta_3\lambda}_{\sigma\rho}+cycle(\beta_2,\beta_3,\lambda)=0.
\end{eqnarray}
These conditions are satisfied due to the Jacobi identities
(\ref{n2}) -- (\ref{n4}) and the restrictions (\ref{m8}) (see
Appendix B). All terms appearing in the equations (\ref{n19}) are
totally antisymmetric in indices $\beta_1,\beta_2,\beta_3$ and
repeat the symmetry properties of
$U^{(3)\beta_1\beta_2\beta_3}_{\;\;\alpha_1\alpha_2\alpha_3\alpha_4}$.
We point out that the restrictions (\ref{m8}) lead to equalities
\begin{eqnarray}
\label{n22}
\bar{V}^{\sigma\beta_1}_{[\alpha_1\alpha_2}
\bar{V}^{\beta_2\rho}_{\alpha_3\alpha_4]}
\bar{V}^{\beta_3\lambda}_{\sigma\rho}=0,\quad
\tilde{V}^{\sigma\beta_1}_{[\alpha_1\alpha_2}
\tilde{V}^{\beta_2\rho}_{\alpha_3\alpha_4]}
\bar{V}^{\beta_3\lambda}_{\sigma\rho}=0,\quad
\bar{V}^{\sigma[\beta_1}_{[\alpha_1\alpha_2}
\tilde{V}^{\beta_2]\rho}_{\alpha_3\alpha_4]}
\bar{V}^{\beta_3\lambda}_{\sigma\rho}=0.
\end{eqnarray}
Therefore we find the solution for structure functions $U^{(3)}$
\begin{eqnarray}
\label{n222}
U^{(3)\beta_1\beta_2\beta_3}_{\;\;\alpha_1\alpha_2\alpha_3\alpha_4}=
-\frac{1}{144}\Big(\bar{V}^{\sigma\beta_1}_{[\alpha_1\alpha_2}
\bar{V}^{\beta_2\rho}_{\alpha_3\alpha_4]}F^{\beta_3}_{\sigma\rho}+
\bar{V}^{\sigma[\beta_1}_{[\alpha_1\alpha_2}
\tilde{V}^{\beta_2]\rho}_{\alpha_3\alpha_4]}F^{\beta_3}_{\sigma\rho}+
\tilde{V}^{\sigma\beta_1}_{[\alpha_1\alpha_2}
\tilde{V}^{\beta_2\rho}_{\alpha_3\alpha_4]}F^{\beta_3}_{\sigma\rho}\Big)
\end{eqnarray}
and the BRST charge in the forth order
\begin{eqnarray}
\label{n23}
{\cal Q}_4=-\frac{1}{24}c^{\alpha_1}c^{\alpha_2}c^{\alpha_3}c^{\alpha_4}
\Big(\bar{V}^{\sigma\beta_1}_{\alpha_1\alpha_2}
\bar{V}^{\beta_2\rho}_{\alpha_3\alpha_4}F^{\beta_3}_{\sigma\rho}+
2\bar{V}^{\sigma\beta_1}_{\alpha_1\alpha_2}
\tilde{V}^{\beta_2\rho}_{\alpha_3\alpha_4}F^{\beta_3}_{\sigma\rho}+
\tilde{V}^{\sigma\beta_1}_{\alpha_1\alpha_2}
\tilde{V}^{\beta_2\rho}_{\alpha_3\alpha_4}F^{\beta_3}_{\sigma\rho}\Big)
{\cal P}_{\beta_1}{\cal P}_{\beta_2}{\cal P}_{\beta_3}.
\end{eqnarray}
If we additionally assume the following restrictions on the
structure constants
\begin{eqnarray}
\label{m9}
&&V^{(k)\beta\beta_1\sigma_1...\sigma_{k-1}}_{[\alpha_1\alpha_2}
V^{(m-k)\beta_2\gamma\sigma_k...\sigma_{m-2}}_{\alpha_3\alpha_4]}
F^{\beta_3}_{\beta\gamma}=0,\\
\nonumber
&&k=1,...,n-1,\quad m>k,\quad m=2,...,2n-2,
\end{eqnarray}
then we have
\begin{eqnarray}
\label{n25}
\bar{V}^{\sigma\beta_1}_{[\alpha_1\alpha_2}
\bar{V}^{\beta_2\rho}_{\alpha_3\alpha_4]}F^{\beta_3}_{\sigma\rho}=0,\quad
\bar{V}^{\sigma[\beta_1}_{[\alpha_1\alpha_2}
\tilde{V}^{\beta_2]\rho}_{\alpha_3\alpha_4]}F^{\beta_3}_{\sigma\rho}=0,\quad
\tilde{V}^{\sigma\beta_1}_{[\alpha_1\alpha_2}
\tilde{V}^{\beta_2\rho}_{\alpha_3\alpha_4]}F^{\beta_3}_{\sigma\rho}=0
\end{eqnarray}
and as the result
\begin{eqnarray}
\label{n26}
U^{(3)\beta_1\beta_2\beta_3}_{\;\;\alpha_1\alpha_2\alpha_3\alpha_4}=0,\quad
{\cal Q}_4=0.
\end{eqnarray}

Therefore as in case of quadratically algebras, for nonlinear
algebras (\ref{n1}) there exists a unique form of the nilpotent BRST
charge
\begin{eqnarray}\label{n24}
{\cal Q}=c^{\alpha}T_{\alpha}-\frac{1}{2}c^{\alpha}c^{\beta}
\Big(F^{\;\;\gamma}_{\alpha\beta}+
{\bar V}^{\gamma\beta}_{\alpha\beta}T_{\beta}\Big){\cal P}_{\gamma}
\end{eqnarray}
if conditions (\ref{m8}), (\ref{m9}) are fulfilled.

We point out, that the universal and  simple enough BRST charge
(\ref{n24}) for nonlinear algebras (\ref{n1}) is found only under
the special conditions (\ref{m8}), (\ref{m9}) on structure of the
algebras. Although these conditions look like very restrictive,
there exist the interesting algebras where they are fulfilled. For
example, the conditions (\ref{m8}), (\ref{m9}) take place for
Zamolodchikov's $W_{3}$ algebra with central extension
\cite{z}\footnote{This algebra is also called quantum spin 3 algebra
\cite{SSvN}.} and for the higher spin algebras in AdS space
\cite{HSF}. Of course, there exist non-linear algebras for which the
conditions (\ref{m8}) and (\ref{m9}) are not fulfilled, e.g. these
relations are not valid for $so(N)$-extended superconformal algebras
with central extension \cite{k} (see also \cite{SSvN}).

\section{Summary}

In this paper we have studied a construction of the nilpotent
classical BRST charge for nonlinear algebras of the form (\ref{n1})
which are characterized by the structure constants
$F^{\;\;\gamma}_{\alpha\beta},V^{(1)\alpha_1\alpha_2}_{\alpha\beta},...,
V^{(n-1)\alpha_1...\alpha_n}_{\alpha\beta}$. The results obtained
are formulated as follows:

{\bf a.} For quadratically nonlinear algebras the explicit form of
the BRST charge in the third order is found without any additional
restrictions on structure constants.

{\bf b.} In the case when the structure constants satisfies the
restrictions (\ref{m8}) the BRST charge is constructed up to the
forth order in the ghost fields.

{\bf c.} If the conditions (\ref{m8}) and (\ref{m9}) are satisfied
and a set of constraints $T_{\alpha}$ is linearly independent, the
BRST charge is given in the universal form (\ref{n24}).

Also we have proved that suitable quantities in terms of which one
can efficiently analyze general nonlinear algebras (\ref{n1}) are
${\bar V}^{\mu\nu}_{\alpha\beta},{\tilde V}^{\mu\nu}_{\alpha\beta}$.

\section*{Acknowledgements}
The authors are grateful to I.V. Tyutin for discussions. The work was partially
supported by
the INTAS grant, project INTAS-03-51-6346, the RFBR grant, project
No.\ 06-02-16346, grant for LRSS, project No.\ 4489.2006.2, the DFG grant,
project No.\ 436 RUS 113/669/0-3 and joint RFBR-DFG grant, project No.\
06-02-04012.


\vspace{.5cm}
\noindent{\Large{\bf Appendix}}
\hspace*{\parindent}
\vspace{0.5cm}
\begin{appendix}

\section{Solutions to equations (\ref{q17}) in the simplest case}
\renewcommand{\theequation}{\thesection.\arabic{equation}}
\setcounter{equation}{0}
\hspace*{\parindent}
\vspace{0.5cm}
\label{APP_A}

To support our proposal concerning the structure of solutions to
Eqs. (\ref{q17}) as well as the arbitrariness in these solutions, we
solve the equations
\begin{eqnarray}\label{a1}
K^{[\alpha\beta]\sigma}_{\alpha_1\alpha_2\alpha_3}+
K^{[\alpha\sigma]\beta}_{\alpha_1\alpha_2\alpha_3}=
-V^{\gamma\alpha}_{[\alpha_1\alpha_2}V^{\beta\sigma}_{\alpha_3]\gamma}.
\end{eqnarray}
together with the Jacobi identities
\begin{eqnarray}\label{a2}
V^{\gamma\alpha}_{[\alpha_1\alpha_2}V^{\beta\sigma}_{\alpha_3]\gamma}+
V^{\gamma\beta}_{[\alpha_1\alpha_2}V^{\sigma\alpha}_{\alpha_3]\gamma}+
V^{\gamma\sigma}_{[\alpha_1\alpha_2}V^{\alpha\beta}_{\alpha_3]\gamma}=0
\end{eqnarray}
in the simplest case when there are only three constraints ($\alpha = 1,2,3$)
in the algebra
(\ref{q1}). To simplify notations we omit lower indices
$[\alpha_1\alpha_2\alpha_3]$ and introduce quantities
\begin{eqnarray}\label{a3}
K^{[\alpha\beta]\sigma}\equiv
K^{[\alpha\beta]\sigma}_{\alpha_1\alpha_2\alpha_3},\quad
R^{\alpha(\beta\sigma)}\equiv -
V^{\gamma\alpha}_{[\alpha_1\alpha_2}V^{\beta\sigma}_{\alpha_3]\gamma}.
\end{eqnarray}
In terms of $R^{\alpha(\beta\sigma)}$ the Jacobi identities (\ref{a2}) are
rewritten as
\begin{eqnarray}\label{a4}
R^{\alpha(\beta\sigma)}+R^{\beta(\sigma\alpha)}+R^{\sigma(\alpha\beta)}=0.
\end{eqnarray}
From (\ref{a1}) it follows the complete set of equations
\begin{eqnarray}\label{a5}
\nonumber
&&K^{[12]1}=R^{1(12)},\quad K^{[21]1}=\frac{1}{2}R^{2(11)}, \quad
K^{[31]1}=\frac{1}{2}R^{3(11)},\quad K^{[13]1}=R^{1(13)},\\
\nonumber
&&K^{[12]2}=\frac{1}{2}R^{1(22)},\quad K^{[21]2}=R^{2(12)}, \quad
K^{[13]3}=\frac{1}{2}R^{1(33)},\quad K^{[31]3}=R^{3(13)},
\\
&&K^{[23]2}=R^{2(23)},\quad K^{[32]2}=\frac{1}{2}R^{3(22)},
 \quad K^{[32]3}=R^{3(23)},\quad
 K^{[23]3}=\frac{1}{2}R^{2(33)},
\\
\nonumber
&& K^{[21]3}+K^{[23]1}=R^{2(13)}, \quad K^{[31]2}+K^{[32]1}=R^{3(12)},
\quad K^{[12]3}+K^{[13]2}=R^{1(23)}.
\end{eqnarray}
Consider equations for $K^{[12]1}$ and $K^{[21]1}$. Due to the
Jacobi identities (\ref{a4}) we have $2R^{1(12)}=-R^{2(11)}$ and
therefore the required symmetry property  $K^{[12]1}=-K^{[21]1}$ is
fulfilled as the consequence of solution to Eqs. (\ref{a5}).
Moreover we have the following representation of $K^{[12]1}$
\begin{eqnarray}\label{a6}
K^{[12]1}=\frac{1}{3}\Big(R^{1(21)}-R^{2(11)}\Big).
\end{eqnarray}
In a similar way, we obtain
\begin{eqnarray}\label{a7}
\nonumber
&&K^{[13]1}=\frac{1}{3}\Big(R^{1(31)}-R^{3(11)}\Big),\quad
K^{[21]2}=\frac{1}{3}\Big(R^{2(12)}-R^{1(22)}\Big),\quad \\
&&K^{[31]3}=\frac{1}{3}\Big(R^{3(13)}-R^{1(33)}\Big),\quad
K^{[23]2}=\frac{1}{3}\Big(R^{2(32)}-R^{3(22)}\Big),\\
\nonumber
&&K^{[32]3}=\frac{1}{3}\Big(R^{3(23)}-R^{2(33)}\Big).
\end{eqnarray}
For remaining quantities we propose the following form
\begin{eqnarray}\label{a8}
\nonumber
&&K^{[12]3}=\frac{1}{3}\Big(R^{1(23)}-R^{2(13)}\Big)+X_1,\quad
K^{[21]3}=\frac{1}{3}\Big(R^{2(13)}-R^{1(23)}\Big)+X_2,\quad \\
&&K^{[13]2}=\frac{1}{3}\Big(R^{1(32)}-R^{3(12)}\Big)+X_3,\quad
K^{[31]2}=\frac{1}{3}\Big(R^{3(12)}-R^{1(32)}\Big)+X_4,\quad \\
\nonumber
&&K^{[23]1}=\frac{1}{3}\Big(R^{2(31)}-R^{3(21)}\Big)+X_5,\quad
K^{[32]1}=\frac{1}{3}\Big(R^{3(21)}-R^{2(31)}\Big)+X_6.
\end{eqnarray}
Substituting these quantities into the last line of (\ref{a5}) and
using the Jacobi identities (\ref{a4}), we obtain
\begin{eqnarray}\label{a9}
X_1+X_3=0,\quad X_2+X_5=0,\quad  X_4+X_6=0.
\end{eqnarray}
The required symmetry properties of $K^{[12]3},K^{[13]2}$ and $K^{[23]1}$
lead to relations
\begin{eqnarray}\label{a10}
X_1=-X_2,\quad X_3=-X_4,\quad  X_5=-X_6.
\end{eqnarray}
Therefore
\begin{eqnarray}\label{a11}
X_1=-X_2=-X_3=X_4= X_5=-X_6
\end{eqnarray}
and the general solution to Eqs. (\ref{a5}) is characterized by the only
constant $X_1\equiv X^{[123]}$ and can be presented in the form
\begin{eqnarray}\label{a12}
K^{[\alpha\beta]\gamma}=\frac{1}{3}\Big(R^{\alpha(\beta\gamma)}
-R^{\beta(\alpha\gamma)}\Big)+X^{[\alpha\beta\gamma]}.
\end{eqnarray}
Returning to equations (\ref{a1}) we find the general solutions in the form
\begin{eqnarray}\label{a13}
K^{[\alpha\beta]\sigma}_{[\alpha_1\alpha_2\alpha_3]}=
-\frac{1}{3}
V^{\gamma[\alpha}_{[\alpha_1\alpha_2}V^{\beta]\sigma}_{\alpha_3]\gamma}+
X^{[\alpha\beta\sigma]}_{[\alpha_1\alpha_2\alpha_3]}
\end{eqnarray}
proposed in Section 2.

\section{Proof of identities (\ref{3rdR2}), (\ref{n20}), (\ref{n21})
 and (\ref{n22})}
\label{APP_B}
\renewcommand{\theequation}{\thesection.\arabic{equation}}
\setcounter{equation}{0}
\hspace*{\parindent}
\vspace{0.5cm}

In this Appendix we prove identities used in Sections 2 and 3. Let
us start with (\ref{3rdR2}). Consider the Jacobi identities
(\ref{q4}) written in the form
\begin{eqnarray}\label{B1}
F^{\gamma}_{\alpha_3\alpha_4}V^{\beta_2\beta_3}_{\beta\gamma}+
F^{\gamma}_{\alpha_4\beta}V^{\beta_2\beta_3}_{\alpha_3\gamma}+
F^{\gamma}_{\beta\alpha_3}V^{\beta_2\beta_3}_{\alpha_4\gamma}+
V^{\gamma(\beta_2}_{\alpha_3\alpha_4}F^{\beta_3)}_{\beta\gamma}+
V^{\gamma(\beta_2}_{\alpha_4\beta}F^{\beta_3)}_{\alpha_3\gamma}+
V^{\gamma(\beta_2}_{\beta\alpha_3}F^{\beta_3)}_{\alpha_4\gamma}=0.
\end{eqnarray}
Multiplying these identities by $V^{\beta\beta_1}_{\alpha_1\alpha_2}$ and
summarizing over $\beta$ we have
\begin{eqnarray}\label{B2}\nonumber
&&V^{\beta\beta_1}_{\alpha_1\alpha_2}
F^{\gamma}_{\alpha_3\alpha_4}V^{\beta_2\beta_3}_{\beta\gamma}+
V^{\beta\beta_1}_{\alpha_1\alpha_2}
F^{\gamma}_{\alpha_4\beta}V^{\beta_2\beta_3}_{\alpha_3\gamma}+
V^{\beta\beta_1}_{\alpha_1\alpha_2}
F^{\gamma}_{\beta\alpha_3}V^{\beta_2\beta_3}_{\alpha_4\gamma}+\\
&&V^{\beta\beta_1}_{\alpha_1\alpha_2}
V^{\gamma(\beta_2}_{\alpha_3\alpha_4}F^{\beta_3)}_{\beta\gamma}+
V^{\beta\beta_1}_{\alpha_1\alpha_2}
V^{\gamma(\beta_2}_{\alpha_4\beta}F^{\beta_3)}_{\alpha_3\gamma}+
V^{\beta\beta_1}_{\alpha_1\alpha_2}
V^{\gamma(\beta_2}_{\beta\alpha_3}F^{\beta_3)}_{\alpha_4\gamma}=0.
\end{eqnarray}
Antisymmetrizing these identities in indices
$\alpha_1,\alpha_2,\alpha_3,\alpha_4$
we obtain the relations
\begin{eqnarray}\label{B3}\nonumber
V^{\beta\beta_1}_{[\alpha_1\alpha_2}
F^{\gamma}_{\alpha_3\alpha_4]}V^{\beta_2\beta_3}_{\beta\gamma}&-&
V^{\beta\beta_1}_{[\alpha_1\alpha_2}
F^{\gamma}_{\alpha_3]\beta}V^{\beta_2\beta_3}_{\alpha_4\gamma}+
V^{\beta\beta_1}_{[\alpha_1\alpha_2}
F^{\gamma}_{\alpha_4]\beta}V^{\beta_2\beta_3}_{\alpha_3\gamma}+\\
\nonumber
&&V^{\beta\beta_1}_{[\alpha_1\alpha_4}
F^{\gamma}_{\alpha_3]\beta}V^{\beta_2\beta_3}_{\alpha_2\gamma}+
V^{\beta\beta_1}_{[\alpha_2\alpha_3}
F^{\gamma}_{\alpha_4]\beta}V^{\beta_2\beta_3}_{\alpha_1\gamma}+
\\
\nonumber
V^{\beta\beta_1}_{[\alpha_1\alpha_2}
V^{\gamma(\beta_2}_{\alpha_3\alpha_4]}F^{\beta_3)}_{\beta\gamma}&-&
V^{\beta\beta_1}_{[\alpha_1\alpha_2}
V^{\gamma(\beta_2}_{\alpha_3]\beta}F^{\beta_3)}_{\alpha_4\gamma}+
V^{\beta\beta_1}_{[\alpha_1\alpha_2}
V^{\gamma(\beta_2}_{\alpha_4]\beta}F^{\beta_3)}_{\alpha_3\gamma}+\\
&&V^{\beta\beta_1}_{[\alpha_1\alpha_4}
V^{\gamma(\beta_2}_{\alpha_3]\beta}F^{\beta_3)}_{\alpha_2\gamma}+
V^{\beta\beta_1}_{[\alpha_2\alpha_3}
V^{\gamma(\beta_2}_{\alpha_4]\beta}F^{\beta_3)}_{\alpha_1\gamma}
=0.
\end{eqnarray}
Note that four last terms are equal to zero because of (\ref{R1}).
Using the Jacobi identities (\ref{q4}) we can rewrite (\ref{B3}) as
follows
\begin{eqnarray}\label{B4}\nonumber
&&V^{\beta\beta_1}_{[\alpha_1\alpha_2}
F^{\gamma}_{\alpha_3\alpha_4]}V^{\beta_2\beta_3}_{\beta\gamma}+
V^{\beta\beta_1}_{[\alpha_1\alpha_2}
V^{\gamma(\beta_2}_{\alpha_3\alpha_4]}F^{\beta_3)}_{\beta\gamma}+\\
\nonumber
&&
V^{\beta\gamma}_{[\alpha_1\alpha_2}
F^{\beta_1}_{\alpha_3]\beta}V^{\beta_2\beta_3}_{\alpha_4\gamma}-
V^{\beta\gamma}_{[\alpha_1\alpha_2}
F^{\beta_1}_{\alpha_4]\beta}V^{\beta_2\beta_3}_{\alpha_3\gamma}-
V^{\beta\gamma}_{[\alpha_1\alpha_4}
F^{\beta_1}_{\alpha_3]\beta}V^{\beta_2\beta_3}_{\alpha_2\gamma}-
V^{\beta\gamma}_{[\alpha_2\alpha_3}
F^{\beta_1}_{\alpha_4]\beta}V^{\beta_2\beta_3}_{\alpha_1\gamma}+
\\
&&
F^{\beta}_{[\alpha_1\alpha_2}V^{\gamma\beta_1}_{\alpha_3]\beta}
V^{\beta_2\beta_3}_{\alpha_4\gamma}-
F^{\beta}_{[\alpha_1\alpha_2}V^{\gamma\beta_1}_{\alpha_4]\beta}
V^{\beta_2\beta_3}_{\alpha_3\gamma}-
F^{\beta}_{[\alpha_1\alpha_4}V^{\gamma\beta_1}_{\alpha_3]\beta}
V^{\beta_2\beta_3}_{\alpha_2\gamma}-
F^{\beta}_{[\alpha_2\alpha_3}V^{\gamma\beta_1}_{\alpha_4]\beta}
V^{\beta_2\beta_3}_{\alpha_1\gamma}
=0.
\end{eqnarray}
The summand in the second line of (\ref{B4}) is equal to zero because it can be
presented as
\begin{eqnarray}\label{B5}\nonumber
&&V^{\beta\gamma}_{[\alpha_1\alpha_2}
F^{\beta_1}_{\alpha_3]\beta}V^{\beta_2\beta_3}_{\alpha_4\gamma}-
V^{\beta\gamma}_{[\alpha_1\alpha_2}
F^{\beta_1}_{\alpha_4]\beta}V^{\beta_2\beta_3}_{\alpha_3\gamma}-
V^{\beta\gamma}_{[\alpha_1\alpha_4}
F^{\beta_1}_{\alpha_3]\beta}V^{\beta_2\beta_3}_{\alpha_2\gamma}-
V^{\beta\gamma}_{[\alpha_2\alpha_3}
F^{\beta_1}_{\alpha_4]\beta}V^{\beta_2\beta_3}_{\alpha_1\gamma}=\\
&&F^{\beta_1}_{\alpha_1\beta}V^{\beta\gamma}_{[\alpha_2\alpha_3}
V^{\beta_2\beta_3}_{\alpha_4]\gamma}+
F^{\beta_1}_{\alpha_2\beta}V^{\beta\gamma}_{[\alpha_1\alpha_4}
V^{\beta_2\beta_3}_{\alpha_3]\gamma}+
F^{\beta_1}_{\alpha_3\beta}V^{\beta\gamma}_{[\alpha_1\alpha_2}
V^{\beta_2\beta_3}_{\alpha_4]\gamma}+
F^{\beta_1}_{\alpha_4\beta}V^{\beta\gamma}_{[\alpha_1\alpha_3}
V^{\beta_2\beta_3}_{\alpha_2]\gamma}=0
\end{eqnarray}
due to (\ref{R1}). In its turn, taking into account (\ref{R1}),
we obtain the following representation of the summand in the third
line of (\ref{B4})
\begin{eqnarray}\label{B6}\nonumber
F^{\beta}_{[\alpha_1\alpha_2}V^{\gamma\beta_1}_{\alpha_3]\beta}
V^{\beta_2\beta_3}_{\alpha_4\gamma}-
F^{\beta}_{[\alpha_1\alpha_2}V^{\gamma\beta_1}_{\alpha_4]\beta}
V^{\beta_2\beta_3}_{\alpha_3\gamma}&-&
F^{\beta}_{[\alpha_1\alpha_4}V^{\gamma\beta_1}_{\alpha_3]\beta}
V^{\beta_2\beta_3}_{\alpha_2\gamma}-
F^{\beta}_{[\alpha_2\alpha_3}V^{\gamma\beta_1}_{\alpha_4]\beta}
V^{\beta_2\beta_3}_{\alpha_1\gamma}=\\
&=&F^{\beta}_{[\alpha_1\alpha_2}V^{\gamma\beta_1}_{\alpha_3\alpha_4]}
V^{\beta_2\beta_3}_{\beta\gamma}.
\end{eqnarray}
Then, by virtue of the obvious symmetry properties
\begin{eqnarray}\label{B7}
F^{\beta}_{[\alpha_1\alpha_2}V^{\gamma\beta_1}_{\alpha_3\alpha_4]}
V^{\beta_2\beta_3}_{\beta\gamma}=
V^{\gamma\beta_1}_{[\alpha_1\alpha_2}F^{\beta}_{\alpha_3\alpha_4}
V^{\beta_2\beta_3}_{\beta\gamma}=-
V^{\gamma\beta_1}_{[\alpha_1\alpha_2}F^{\beta}_{\alpha_3\alpha_4}
V^{\beta_2\beta_3}_{\gamma\beta},
\end{eqnarray}
from (\ref{B4}) we finally find
\begin{eqnarray}\label{B8}
V^{\beta\beta_1}_{[\alpha_1\alpha_2}
V^{\gamma(\beta_2}_{\alpha_3\alpha_4]}F^{\beta_3)}_{\beta\gamma}=0.
\end{eqnarray}
which is the first identities in (\ref{3rdR2}).

To prove the second identities in (\ref{3rdR2}), consider
 the Jacobi identities (\ref{q5}) written in the form
\begin{eqnarray}\label{b1}
V^{\gamma(\beta_2}_{\alpha_3\alpha_4}V^{\beta_3\sigma)}_{\beta\gamma}+
V^{\gamma(\beta_2}_{\alpha_4\beta}V^{\beta_3\sigma)}_{\alpha_3\gamma}+
V^{\gamma(\beta_2}_{\beta\alpha_3}V^{\beta_3\sigma)}_{\alpha_4\gamma}=0.
\end{eqnarray}
Multiplying these identities by
$V^{\beta\beta_1}_{\alpha_1\alpha_2}$ and summing over $\beta$ we
have
\begin{eqnarray}\label{b2}
V^{\beta\beta_1}_{\alpha_1\alpha_2}
V^{\gamma(\beta_2}_{\alpha_3\alpha_4}V^{\beta_3\sigma)}_{\beta\gamma}+
V^{\beta\beta_1}_{\alpha_1\alpha_2}
V^{\gamma(\beta_2}_{\alpha_4\beta}V^{\beta_3\sigma)}_{\alpha_3\gamma}+
V^{\beta\beta_1}_{\alpha_1\alpha_2}
V^{\gamma(\beta_2}_{\beta\alpha_3}V^{\beta_3\sigma)}_{\alpha_4\gamma}=0.
\end{eqnarray}
After antisymmetrizing in indices $\alpha_1,\alpha_2,\alpha_3,\alpha_4$
we derive
\begin{eqnarray}\label{b3}\nonumber
V^{\beta\beta_1}_{[\alpha_1\alpha_2}
V^{\gamma(\beta_2}_{\alpha_3\alpha_4]}V^{\beta_3\sigma)}_{\beta\gamma}
&=&
V^{\beta\beta_1}_{[\alpha_1\alpha_2}
V^{\gamma(\beta_2}_{\alpha_3]\beta}V^{\beta_3\sigma)}_{\alpha_4\gamma}-
V^{\beta\beta_1}_{[\alpha_4\alpha_1}
V^{\gamma(\beta_2}_{\alpha_2]\beta}V^{\beta_3\sigma)}_{\alpha_3\gamma}+\\
&&
V^{\beta\beta_1}_{[\alpha_3\alpha_4}
V^{\gamma(\beta_2}_{\alpha_1]\beta}V^{\beta_3\sigma)}_{\alpha_2\gamma}-
V^{\beta\beta_1}_{[\alpha_2\alpha_3}
V^{\gamma(\beta_2}_{\alpha_4]\beta}V^{\beta_3\sigma)}_{\alpha_1\gamma}.
\end{eqnarray}
Each term in rhs. of (\ref{b3}) is equal to zero because of
(\ref{R1}). Therefore we obtain
\begin{eqnarray}\label{b4}
V^{\beta\beta_1}_{[\alpha_1\alpha_2}
V^{\gamma(\beta_2}_{\alpha_3\alpha_4]}V^{\beta_3\sigma)}_{\beta\gamma}
=0.
\end{eqnarray}
If we start with the relations (\ref{R1}) then in a similar way we
can obtain the relations
\begin{eqnarray}\label{b5}
V^{\beta\beta_1}_{[\alpha_1\alpha_2}
V^{\gamma\beta_2}_{\alpha_3\alpha_4]}V^{\beta_3\sigma}_{\beta\gamma}
=0
\end{eqnarray}
which were used in Section 2 to find structure functions
$U^{(3)\beta_1\beta_2\beta_3}_{\alpha_1\alpha_2\alpha_3\alpha_4}$.

Now we consider the proof of identities (\ref{n20}) and (\ref{n21}).
In the case when restrictions (\ref{m8}) are valid then the Jacobi
identities (\ref{n}) and (\ref{n3}) are reduced to
\begin{eqnarray}\label{b6}
F^{\;\;\rho}_{[\alpha\beta}V^{(m)\beta_1...\beta_m\beta_{m+1}}_{\lambda]\rho}
+ V^{(m)\rho(\beta_1...\beta_m}_{[\alpha\beta}
F^{\;\;\beta_{m+1})}_{\lambda]\rho}=0, \quad  m=1,...,n-1.
\end{eqnarray}
From (\ref{m8}) and (\ref{b6}) one can easy derive the following relations
\begin{eqnarray}\label{b7}
{\bar V}^{\sigma\alpha}_{[\mu\nu}{\bar V}^{\beta\gamma}_{\lambda]\sigma}=0,\quad
{\tilde  V}^{\sigma\alpha}_{[\mu\nu}{\tilde V}^{\beta\gamma}_{\lambda]\sigma}=0,
\quad
{\bar V}^{\sigma\alpha}_{[\mu\nu}{\tilde V}^{\beta\gamma}_{\lambda]\sigma}=0,
\quad
{\tilde V}^{\sigma\alpha}_{[\mu\nu}{\bar V}^{\beta\gamma}_{\lambda]\sigma}=0
\end{eqnarray}
and
\begin{eqnarray}\label{b8}
F^{\;\;\rho}_{[\alpha\beta}{\bar V}^{\beta_1\beta_2}_{\lambda]\rho}
+ {\bar V}^{\rho(\beta_1}_{[\alpha\beta}
F^{\;\;\beta_2)}_{\lambda]\rho}=0, \quad
F^{\;\;\rho}_{[\alpha\beta}{\tilde V}^{\beta_1\beta_2}_{\lambda]\rho}
+ {\tilde V}^{\rho(\beta_1}_{[\alpha\beta}
F^{\;\;\beta_2)}_{\lambda]\rho}=0
\end{eqnarray}
respectively.

Taking into account the proof given in the beginning of this
Appendix one can conclude that from (\ref{b7}) and (\ref{b8}) it
follows that
\begin{eqnarray}\label{b9}
\bar{V}^{\sigma\beta_1}_{[\alpha_1\alpha_2}
\bar{V}^{\rho(\beta_2}_{\alpha_3\alpha_4]}F^{\beta_3)}_{\sigma\rho}=0,\quad
\tilde{V}^{\sigma\beta_1}_{[\alpha_1\alpha_2}
\tilde{V}^{\rho(\beta_2}_{\alpha_3\alpha_4]}F^{\beta_3)}_{\sigma\rho}=0,\quad
\bar{V}^{\sigma\beta_1}_{[\alpha_1\alpha_2}
\tilde{V}^{\rho(\beta_2}_{\alpha_3\alpha_4]}F^{\beta_3)}_{\sigma\rho}=0,\\
\label{b10}
\bar{V}^{\sigma\beta_1}_{[\alpha_1\alpha_2}
\bar{V}^{\rho\beta_2}_{\alpha_3\alpha_4]}
\bar{V}^{\beta_3\lambda}_{\sigma\rho}=0,\quad
\tilde{V}^{\sigma\beta_1}_{[\alpha_1\alpha_2}
\tilde{V}^{\rho\beta_2}_{\alpha_3\alpha_4]}
\bar{V}^{\beta_3\lambda}_{\sigma\rho}=0,\quad
\bar{V}^{\sigma\beta_1}_{[\alpha_1\alpha_2}
\tilde{V}^{\beta_2\rho}_{\alpha_3\alpha_4]}
\bar{V}^{\beta_3\lambda}_{\sigma\rho}=0.
\end{eqnarray}
which immediately lead to identities (\ref{n20}), (\ref{n21}) and (\ref{n22}).

\end{appendix}

\end{document}